\documentstyle[11pt,aaspp4]{article}

\newcommand{\beq}{\begin{equation}}
\newcommand{\eeq}{\end{equation}}

\begin{document}

\title{A Study of the $B-V$ Colour Temperature Relation}

\author{Maki Sekiguchi\altaffilmark{1} and 
        Masataka Fukugita\altaffilmark{2,1}}

\affil{$^1$Institute for Cosmic Ray Research, University of Tokyo,
Tanashi, Tokyo 188, Japan}

\affil{$^2$Institute for Advanced Study, Princeton, NJ 08540, USA}

\begin{abstract}
We attempt to construct a $B-V$ colour temperature relation for
stars in the least model dependent way employing the best modern data.
The fit we obtained with the form 
$T_{\rm eff}= T_{\rm eff}((B-V)_0, [{\rm Fe/H}],
\log g)$ is well constrained and a number of tests show the consistency
of the procedures for the fit. Our relation covers from F0 to K5 stars
with metallicity [Fe/H]=$-$1.5 to +0.3 for both dwarfs and giants. The residual
of the fit is 66 K, which is consistent with what are expected from the
quality of the present data. Metallicity and surface gravity effects are
well separated from the colour dependence.
Dwarfs and giants match well in a single family
of fit, differing only in $\log g$. The fit also detects the Galactic
extinction correction for nearby stars with the amount 
$E(B-V)= 0.26\pm0.03$ mag/kpc.
Taking the newly obtained relation as 
a reference we examine
a number of $B-V$ colour temperature relations and atmosphere models 
available in the literature.
We show the presence of a systematic error in the colour temperature 
relation from 
synthetic calculations of model atmospheres; the systematic
error across K0 to K5 dwarfs
is 0.04$-$0.05 mag in $B-V$, which means 0.25-0.3 mag in $M_V$
for the K star range. We also argue for the error in the temperature scale
used in currently popular stellar population synthesis models;
synthetic colours from these models are somewhat too blue for aged
elliptical galaxies. 
We derive the colour index of the sun $(B-V)_\odot=0.627\pm
0.018$, and discuss that redder colours (e.g., 0.66-0.67) often quoted
in the literature are incompatible with the colour-temperature relation
for normal stars.  
\end{abstract}

 
\newpage
\section{Introduction}

The determination of the stellar locus in the HR diagram is a subject
of the prime importance in astrophysics, as well as it has wide applications. 
For instance, the determination of the distance scale relies much on the 
uniqueness of the stellar locus. Such work has often resorts to the
knowledge of theoretical isochrones, since the observations alone do 
not span sufficiently large parameter space. On the other
hand, theoretical isochrones, expressed in the colour-magnitude space, 
may suffer from errors of three origins. The errors may arise from (i)
evolution track calculations which depend on opacity, nuclear reaction rate,
equation of states and the treatment of convection, (ii) conversion 
of temperature to colour
index (colour-temperature relation), and 
(iii) conversion of luminosity to magnitude in a specific passband
(bolometric corrections). The recent findings of the difference in the distance
to open clusters by Hipparcos parallax as compared to the traditional
zero age main sequence (ZAMS) fitting (van Leeuwen \& Hansen Luiz 1997; 
van Leeuwen 1999; Mermilliod et al. 1997; 
Pinsonneault et al. 1998) have tempted us to study the problem
of reliability concerning theoretical isochrones. An analysis of
Nordstr\"om, Andersen \& Andersen (1997) indicates that the discrepancy
of the isochrones among different authors can be as much as 0.4$-$0.6 mag,
although in most practical applications the isochrones are used so that such
a large error does not directly affect the results.
Their figures also indicate that a dominant part
of the errors may arise from the colour-temperature relation, especially when
$B-V$ colour is used, as we have also confirmed from our own analysis. The
error arising from the bolometric correction is rather small, and the
evolution track on the luminosity temperature plane is reasonably
converged among authors, in so far as we are concerned with the region
far from the turn-off point where convective overshooting may start making a
difference.
 
Another particularly important application of the stellar track is 
stellar population synthesis of galaxy colours, which play an
important role in cosmology (e.g., Tinsley \& Gunn
1976; Bruzual \& Charlot 1993; Kodama \& Arimoto 1997).  If colour of
giant stars would contain errors as much as $\Delta (B-V)>0.05$, the
interpretation of elliptical galaxies could significantly be disturbed.

In this paper we focus on the problem of the colour-temperature 
($T_{\rm eff}$) relation. We attempt to construct the relation, which 
we think the least model dependent, and study what errors
are contained in the existing relations. There are a lot of work for the
colour-temperature relation. The early authority is the one given by
Johnson (1966), and the work to 1980 is summarised by B\"ohm-Vitense 
(1980). We also quote several representative examples, which include
Code et al. (1976), Blackwell \& Shallis (1977), Bessell (1979), 
Ridgway et al. (1980), Saxner \& Hammarba\"ck 
(1985), Arribas \& Mart\'inez-Roger (1988), Tsuji et al (1995) and  
Di Benedetto \& Rabbia (1987). 
Most recent work includes  Flower (1996), Alonso et al. (1996b),
Blackwell \& Lynas-Gray (1998; hereafter BL98), 
and Lejeune, Cuisinier \& Buser (1998). 

The prime difficulty lies in estimating precise temperature. The most direct 
method employs the measurement of angular diameter of stars using
interferometry or lunar occultations. The number of stars which are
given accurate angular diameters are increasing (Davis 1998), 
especially with the advancement of the Michelson interferometry technique,
but not yet sufficiently many to explore large parameter space. 

One of the methods which are supposed to be accurate and often used 
in modern literature is the infrared flux 
method (IRFM), in which $T_{\rm eff}$ is estimated from the measurement 
of $R=F_{\rm bol}/F_\lambda=\sigma T_{\rm eff}^4/\psi(\lambda)$ for 
infrared $\lambda$, where  $\psi(\lambda)$ is calculated from model
atmospheres (Blackwell \& Shallis 1977). This method has been 
developed to reduce the model dependence
using the fact that $F_\lambda$ in the near infrared regions is 
smoothly proportional to $T_{\rm eff}$
and model dependence is fairly small. 
Alonso, Arribas \& Mart\'inez-Roger 
(1996a,b) and BL98 
have given the latest and most extensive 
work employing this method.

M\'egessier (1994) and Alonso et al. (1996a)
have examined the error associated with this method.
The authors of both papers claim that the derived temperature differs as
much as 100K depending on the model atmosphere employed in the work.
For example use of ATLAS9 (Kurucz 1993) gives temperature 100K higher than 
ATLAS8. Similar
difference is also reported between MARCS (Gustafsson et al. 1975
and their updates) 
and ATLAS9. 

An important advancement is brought by
Di Benedetto (1998; hereafter B98) 
who found an empirically very tight relationship
between $T_{\rm eff}$ and $V-K$ colour, which is calibrated with
the direct angular diameter measurement. He has shown that this
relationship depends very little on luminosity class and metal abundance.
This makes possible to estimate temperature  for F-K dwarfs, for which 
direct measurements of diameters are still lacking. This method
would offer the least model-dependent  method
to deal with a fairly large sample without a
direct angular size measurement for each star.

We have examined the accuracy of the B98 temperature, and are convicted that
it is perhaps the best method available to us for the time being. 
B98, however, has not discussed much with $B-V$ colour, which is very sensitive
to line blanketing and also to surface gravity, giving a large scatter
around the surface brightness colour relation. This is also true with
a recent comparative study of Bessell, Castelli \& 
Plez (1998), who extensively compared model atmosphere calculations 
with empirically 
estimated $T_{\rm eff}$ for various colour bands, but except for $B-V$.
On the other hand, most of the applications of the stellar
locus still rely predominantly on $V$ magnitude and $B-V$ colour.

For this reason we attempt to construct a $B-V$ colour temperature 
relation adopting the $T_{\rm eff}$ data from B98 together with
an extensive photometric database compiled by Hauck \& Mermilliod (1998),
combined with the metallicity and log $g$ data compiled by  
de Strobel et al. 1997 (hereafter SSFRF) for F0-K7 stars. We also examine the
available $B-V$ colour temperature relations, especially those
published recently,  against what we have 
constructed in order to find external errors of these works. 
Further examination is also made for the colour-temperature relation
used by theoretical work of stellar evolution, which is often used
as a basis to discuss the cosmic distance scale and age, as well as taken as
a fiducial for stellar population synthesis for colour of
galaxies. As for the required accuracy for the distance work,
if our goal is to obtain a 3\% accuracy in the distance estimation, 
one would need to achieve the accuracy of
$B-V$ to be 0.01 mag, which in turn is translated to temperature error of
20$-$45 K. In our paper we attempt to document the error budget arising from 
many components of the input data. This would clarify the limiting factor
to the accuracy of the $B-V$ colour temperature relation and tell us
what improvement should be done to go further.
We shall see that the accuracy that
we can achieve is worse than this goal by about a factor of 2$-$3 for
each star. When many stars in clusters are averaged, however, the accuracy  
is of the order of 0.015 mag in $B-V$.

One of our additional aims is to extract the dependence of the $B-V$ colour 
temperature relation on metallicity, without resorting 
to theoretical grids, and in particular to examine 
the accuracy of synthetic results
from model atmospheres. As a byproduct this makes possible to
estimate colour of the sun from solar 
analogues (e.g., Taylor 1998;
de Strobel 1996 for reviews). 

In section 2, we derive the $B-V$ colour temperature relation, after
examining the quality and reliability of input data. We also present
a table of the estimated error budget. In section 3, an extensive comparison
is made with the existing $B-V$ colour temperature relations based on
either empirical or theoretical ground. We discuss in section 4
$B-V$ colour of the sun. Metallicity dependence is discussed in
section 5. Our conclusion is given in section 6.

\section{Construction of the $B-V$ colour temperature relation}

\subsection{Data}

To derive a relation between $T_{\rm eff}$ and $B-V$ colour, 
we start with 537 ISO standard stars 
for which B98 has given accurate estimates of $T_{\rm eff}$. 
Of those 537 starts, 270 stars 
are given estimates of [Fe/H] and $\log g$ by SSFRF. Among them 
about 40\% (110 stars) are 
dwarfs or subdwarfs and 60\% (160 stars) are giant stars. For a majority
of entries SSFRF gives more than one data for [Fe/H] and $\log g$ 
for a given star;
in such cases we take median values of [Fe/H] and $\log g$.
The distributions of the median values of [Fe/H] and $\log g$ are shown 
in Figs. 1 (a) and (b).  The data are distributed widely from [Fe/H]
=$-$2.0 to +0.5; only 10\% of stars have [Fe/H] $<-1.0$,
but we still have a reasonable number of stars to constrain our analysis
in this region.
The ten percentile for the
metal rich side is [Fe/H] $= 0.12$.
All stars of our sample are given $B-V$ colours (Hauck \& Mermilliod 1998)
and parallaxes with Hipparcos (ESA 1997).

We give in Table 1 the derivative of $T_{\rm eff}$ against $B-V$ to
obtain an idea about the propagation of errors from
$T_{\rm eff}$ to $B-V$. The table gives $\Delta T_{\rm eff}$ that causes
a change of $\Delta (B-V)$=0.01. For instance, $\Delta T_{\rm eff}=45$K
is an allowance for F0 stars (7000K); it is 30 K for G2 stars, and 16K 
for K5 stars at 4000 K.

\subsection{Examination of B98's temperature estimates}
\label{sec_teff}

None of the temperature estimates are completely free from the
atmosphere model. Spectroscopic determinations of temperature
directly rely on the details of the atmosphere model, and
IRFM uses the prediction of the atmosphere model in the near 
infrared region. Even the direct measurement of angular diameters
should be supplemented with the atmosphere model, albeit with 
a minimal extent,
in order to estimate the bolometric flux in the invisible regions.
The method proposed by B98 basically belongs to this last category.
He found a tight correlation between surface brightness and
$V-K$ colour (dispersion being 0.03 mag) to estimate angular size
$\phi$, and used another tight relationship between
the bolometric flux $F_{\rm bol}$ and $V-K$ colour found by BL98 to estimate
effective temperature of stars for which the direct measurements
are not available for $\phi$ and $F_{\rm bol}$.

Since it is of crucial importance to examine the accuracy of 
effective temperature estimated with this method,  
we plot in Fig. 2 the difference of temperature given by B98 from
$V-K$ colours and that obtained directly using 

 \begin{equation}
  T_{\rm eff}^4 = \frac{4}{\sigma} \phi^{-2} F_{\rm bol},
 \label{teff_eq}
 \end{equation}
which is equivalent to the defining equation for effective temperature. Here
$\sigma$ is the Stefan-Boltzmann constant. 
The table given by B98 (Table 3) contains all angular
size measurements which could be used for our test
(21 stars and the sun). We take the bolometric
flux from direct evaluations of several sources, as presented in
Table 2. The bolometric flux is an integration of flux from $U$ to
$K$ bands with shorter and longer wavelength ranges evaluated with
the aid of atmosphere models. The direct integration accounts for 93\% of flux
at $T$=4500K, and 83\% at 7000K (Alonso et al 1995). 
Therefore, the dependence on model
atmosphere is expected to be quite small. It is expected that the error
is no more than 2\% for the bolometric flux, which means
a 0.5\% error in $T_{\rm eff}$. We also note that there is an error
of this order (1.5\%) in the absolute calibration of flux of $\alpha$ Lyr
at 5556\AA~(Hayes 1985). We explicitly document the calibrations employed by
the respective authors in Table 2 (ref/norm), but do not dare to adjust to the 
same scale, since the difference is smaller than errors of other
origins.
Among 22 stars of B98 direct bolometric flux estimates are available for
16 stars. We take each estimate by different authors as an independent data,
so that 32 data points are contained in Fig. 2. The errors attached
to our $T_{\rm eff}$ are obtained by a quadrature of the errors for
$F_{\rm bol}$ and $\phi$ given by the authors. 
We also added the data points from a similar
test of B98 himself (Table 5 of B98) and from 
the $T_{\rm eff}$ estimate of BL98 when
available. 

For stars later than F5 type ($V-K\ge 1.0$), 
errors in $\phi$ and those in  $F_{\rm bol}$
are comparable, and their quadratures ($\delta T_{\rm eff}$) 
are also comparable to the difference between the B98 estimate and our
``true'' (reference) temperature 
($\Delta T_{\rm eff}({\rm B98})=T_{\rm eff}({\rm B98})-T_{\rm eff}({\rm ref)}$).  
We see some systematic
trend that B98 temperature is slightly lower than the reference temperature
by 30$-$40K. We note an excellent agreement between BL98 and B98.
For stars earlier than F0 type the angular size measurement yields too
large errors to carry out an accurate test. 
For all stars formal errors
exceed the difference between of the two temperature estimates.
Except for $\varepsilon$ Sgr
and one point for $\alpha$ Lyr, however, B98 temperature agrees very well
with our reference within $\approx \pm$ 50K, although error bars
are larger. For $\alpha$ Lyr (BS 7001) the two independent flux determinations 
disagree by 4.8\%, which is not explained merely by the different absolute
calibrations they take.
On the other hand, BL98 tend to give temperature
100-200K higher than ours in this region. 

From this test we conclude that B98's estimate
of temperature is correct allowing for errors of 30$-$40K at least
for F0-K8 stars,
($V-K=0.7-3.5$).
We confine ourselves to the range $T_{\rm eff}<7000$ (F0 or
later), for which the error of $T_{\rm eff}$ is small and the result is
reliable to a high accuracy.

\subsection{Errors of $B - V$}
\label{sec_bv}

We take the mean values of $B-V$ compiled by Hauck \& Mermilliod (1998). 
81\% of the stars in our sample have a dispersion of less than 0.01 among
multiple observations, and 93\% of the stars have less than 
0.014 mag dispersion. 
As a conservative estimate we infer the average photometric error 
to be 0.01 mag, which corresponds to 15-45 K in our $T_{\rm eff}$ range. 

Most of stars in our sample are located nearby:
61\% of our stars are within 50 pc,  21\% lie between 50 pc and 100 pc, 
and 11\% between 100 pc and 150 pc.  Therefore, the necessary extinction 
correction
is a minimum amount.
Nevertheless, we apply the extinction correction
$E(B-V)/d=0.235$ mag/kpc, or $A_V/d=0.8$ mag/kpc ($d$ being the distance)
with $R=3.4$, taken from
Blackwell et al. (1990).
We examine the validity of this extinction correction when we obtain
a fit of the form $T_{\rm eff}=T_{\rm eff}((B-V)_0, [{\rm Fe/H]}, \log g)$,
and confirm that extinction correction is indeed indispensable to obtain a good
fit; selective extinction of $E(B-V)/d=0.269$/kpc gives a
minimum to the residuals of the fit (see below). Since the adoption of
this value hardly modifies the results of the fit, we take 0.235 mag/kpc
for our final results and we
take the difference of the two
values evaluated at 100 pc, i.e., 0.0034 mag, as a representative error 
from the extinction correction.

\subsection{Errors of [Fe/H] and $\log g$ estimates}
\label{sec_feh}

The scatter of the [Fe/H] values documented in a catalogue of SSFRF 
is $<0.15$ dex 
for 80\% of our sample, and $<0.2$ dex for 87\% of the sample. Our fit
given below shows a derivative $\partial T_{\rm eff}/\partial$[Fe/H]$\simeq$
320 K/dex. Therefore, the error of the [Fe/H] measurement causes 50 K
in the determination of $T_{\rm eff}$. The uncertainty of $\log g$ does
not cause much errors in $T_{\rm eff}$. The scatter of SSFRF data is
about 0.2 in log $g$ units; the derivative of our fit 
$\partial T_{\rm eff}/\partial \log g=-30$K means the error being about 
6 K. We also expect this order of scatter in $\log g$ from a star to a star
at a given stellar mass (Andersen 1991).

\subsection{Results of the Fit}

We present in Fig. 3 $T_{\rm eff}$ as a function of unreddened $B-V$ colour 
for all 283 stars.  13 out of 16 calibrating stars in Table 2 have 
[Fe/H] and $\log g$ data in SSFRF, and they are also included in the plot.
In this figure we classified stars according to metallicity: solid
circles are stars with [Fe/H]$\le-$0.75, open squares for 
$-0.75<$[Fe/H]$\le-0.25$, cross symbols for $-0.25<$[Fe/H]$\le$+0.25, and 
solid triangles are for  [Fe/H]$>$+0.25. The sample covers the range $0.3\le B-V
\le 1.5$, which is the range of our analysis. 
In Fig. 4  are selected only dwarfs (and sub-dwarfs). In this
sample stars with $B-V>0.9$ are scanty, and we must limit our study to the range
between $B-V=0.3$ and 0.9.

In carrying out our fitting, we exclude 17 stars, 10 of them as having  
$B-V<0.3$, one of them is too distant (700 pc away)
and remaining 6 being located more than 4$\sigma$ away from the
locus of the fit. Our fitting is made in the following steps. First we 
fit the samples 
(full and dwarf samples) simply with $T_{\rm eff}=T_{\rm eff}((B-V)_0)$
ignoring the [Fe/H] and $\log g$ dependence; we find the rms residual
of the fit to be 123K for the full sample and 151 K for the dwarfs. We then 
carry out a fit with the form
\begin{equation}
\log T_{eff} = c_0 + c_1 (B - V)_0 + c_2 (B - V)_0^2 + c_3 (B - V)_0^3 +
                f_1 {\rm [Fe/H]} + f_2 {\rm [Fe/H]}^2 + g_1 \log g.
\label{teff_fit}
\end{equation}
with an equal weight given to all data points. This largely reduces the rms
of residuals. We find 66 K (1.1\%) for the full sample and 71 K (1.2\%)
for dwarfs. This rms is somewhat smaller than that of BL98
(80$-$90 K), and smaller by a factor of two 
than is given by Alonso et al. (1996b) (130 K).  In fitting the dwarf sample,
we fixed $\log g$ to be constant ($\log g=4.3$) (see below), and $g_1$ to the
value derived from the fit to the full sample, since the variation of
$\log g$ for main sequence stars in this colour range is too small
to constrain the fit.  
The parameters of this fit are given in Table 3. Although we have 
a strong correlation among coefficients $c_0-c_3$, and between 
$f_1$ and $f_2$, cross correlations are quite small ($<0.1$ when 
diagonals are normalized
to unity) between $c_i$ and $f_j$. Cross correlations
between $c_i$ and $g_1$, and between $f_i$ and $g_1$ are also
small ($<0.25$ and $<0.15$, respectively).
The fit is well constrained expect for a high temperature range $\log T
>3.82$: the adoption of variables $T$ or $1/T$ in the right hand
of equation (2) does not modify the shape of the curve and the quality of 
the fit, except at the weakly constrained very end of the
high temperature edge where we see a change 
equivalent to $B-V\leq 0.01$.
The parameters for the dwarf sample are consistent with those for the full 
sample, though the former set has larger errors. 
The consistency of the two fits means that the two samples
are well controlled by simply different $\log g$. Hence, we adopt 
the fit with the full sample as our best result and use this for 
further analysis in what follows including that for dwarfs.

We further proceed with our fitting tests. We fit our samples with adding
a cross term [Fe/H] $\times$ (B-V)$_0$. This does not 
reduces rms residuals at all, but merely 
increase the error estimate of the parameters (in particular this
doubles the error of $f_1$), indicating that the cross term is not very
well determined. Actually the cross term thus obtained is rather small, and
it changes $\partial T_{\rm eff}/\partial$ [Fe/H]
only by 14\% between $(B-V)=0.4$ and 0.8.
So we can drop this term. 

The next is a fit ignoring  extinction 
corrections. This raises the residual temperature from 66 K to 73 K, 
indicating the necessity of extinction corrections. 
The significance is demonstrated 
in Fig. 5, where we plot the residual rms as a function of
the selective extinction per unit distance.
We see that the minimum is attained with $E(B-V)=0.269$/kpc, and our
adopted value 0.235 increases a residual only less than 0.2 K
\footnote[1]{Within 50 pc of the solar neighbourhood, the extinction
seems to be somewhat smaller than this value. For the stars within
50 pc, we obtain $0.21\pm0.06$ mag/kpc from a sample of 151 stars.
This means that extinction increases to $\sim0.35$  mag/kpc
at $\approx$ 100pc. We thank Bohdan Paczy\'nski for attracting our
attention to this problem}. 
We obtain an error of 0.030/kpc for this parameter when this is
allowed to vary as a free parameter.
This gives
not only an excellent confirmation of the selective extinction per distance
used in the literature (e.g. Blackwell et al. 1990), but also shows that 
our fit would differentiate such a small changes in the data, indicating 
an overall consistency of the fitting procedures.

The overall quality of our best fit is shown in Fig. 6, where 
the effective temperature data are corrected for metallicity and 
surface gravity according to  
$T_{eff} -( f_1 {\rm [Fe/H]} + f_2 {\rm [Fe/H]}^2 + g_1 \log g )$, 
and those
points plotted are supposed to give data for [Fe/H]=0 and $\log g$=4.3
as a function of 
$(B-V)_0$. Fig. 7 shows residuals in more detail, where we see that
giants and dwarfs are indeed on the same family simply with different
$\log g$. This would justify to use the same family of curves for the
entire range of our colour space. The data points with squares are stars
used for examination of temperature above (we plot only medians when a
number of bolometric flux estimates are available).

\subsection{Error budget}

We have already discussed the source of errors. Our estimate
of the size of errors is summarized in Table 4 for F5, G2 and K5 stars.
The sources we discussed above all contribute to the dispersion of the
final fit. The quadrature of internal error budget amounts to
67$-$77 K, which is consistent with the actual dispersion of
the fit 60$-$80 K (global value is 66 K). 
This means that the error propagation is well controlled
in our data processing procedures, and intrinsic scatter of the
$(B-V)$ colour temperature relation is substantially smaller than  
$\approx$40 K.

We note that the error of [Fe/H] and that of $T_{\rm eff}$ are comparable
and are the dominant source of errors. Photometry error might compete with
this for early type stars, where the curve gets steeper.
Errors from other entries are
smaller. This dispersion of temperature 
corresponds to $\delta (B-V)\simeq 0.02$
and increases to 0.03 for low temperature stars. 
Of course, the locus of the relation is better determined. We anticipate
a systematic error up to about 30-40 K in the B98 temperature estimate,
and the normalisation error of the bolometric flux, which is used
as an external calibrator in our work, of the order of 1.5\%
(0.37\% in temperature), and there may be a systematic trend of
metallicity scale on the order of 0.05 dex depending on authors.
This makes overall systematic error to be about 45 K (these systematic
errors are significantly smaller than the random errors, and are
supposed to be already included in the error budget shown in Table 4).  
This seems to be the best we can achieve with the present data.

\section{Comparison of the $B-V/T_{\rm eff}$ relations 
in the literature}

\subsection{Dwarfs}

We discuss the $B-V$ colour temperature relations 
($B-V/T_{\rm eff}$ relations) available in the
literature, taking the one we obtained above as a reference.
Fig. 8 is a compilation of the colour temperature relations for
the main sequence stars with solar metallicity.
To draw the locus of our colour temperature relation we assume
$\log g=129.34 - 64.66\log T + 8.347(\log T)^2$,
which is obtained by fitting data of SSFRF
for dwarfs used in our analysis. This relation differs from what would
be obtained by fitting the data from binary stars (Popper 1980)
by an amount of $\Delta
\log g\simeq 0.2$ for G stars, but the scatter indicated by the B98
sample (with SSFRF data for $\log g$) and that by the data of Andersen
(1991) are somewhat 
larger than the offset. In any case, the difference caused by
the difference of $\log g$ of this amount is small and it changes the
resulting temperature by no more than $\approx$ 10 K, or $B-V$ by 0.005. 


Now, in order to examine the detail, we display in Fig. 9 the difference 
of various relations against the one
obtained by  Lejeune, Cuisinier \& Buser (1998; hereafter LCB): 
$\Delta(B-V)=(B-V)_0 - (B-V)_{\rm 0,LCB}$.
The adoption of their
relation as a fiducial zero point
is motivated by the fact that their relation covers
the widest $B-V$ range, while the range of the relation we obtained is not
as wide as theirs. In this figure we have plotted 9 relations, which
we are going to discuss in detail: Flower (1996), Alonso et al. (1996b),
BL98, Code et al. (1976), Demarque et al. (1985; Yale isochrone), and
Bertelli et al. (1994), and the relation derived using ATLAS9 (Kurucz 1993)
atmosphere, together with LCB and ours. 
The Yale isochrone and Bertelli et al. are theoretical estimates 
based on model atmospheres, and we have computed colours for
ATLAS9 using the response functions of 
Azusienis and Straizys (1969) for the $B$ and $V$ pass bands.
Table 4 summarizes briefly the methods adopted by the respective authors.
We also plot the position of the sun taking ($T_{\rm eff}=5777\pm6$ K and
$B-V=0.64\pm0.01$, which we discuss in the next section. 

It is clear at a glance that LCB and Flower (1996) are largely
deviated from the others including our newly obtained $B-V/T_{\rm eff}$
relation.
Flower uses temperature information collected from various sources: some 
from direct measurements of angular diameters, some from IRFMs, and 
others from spectroscopic 
analysis. For $\log T_{\rm eff}<3.75$, where a large departure starts, 
Flower's data mostly rely on 
temperature from an IRFM analysis of Bell \& Gustafsson (1989). 
To study a possible problem with Flower's temperature, 
we show in Fig. 10 the stars he used. In the range of our interest
$B-V>0.65$, we find new temperature determinations by  Soubiran, 
Katz \& Cayrel (1998) for 7 stars (indicated by arrows). 
The new temperature for these stars
are significantly lower as we go to redder stars, and they
fall on the curve of  $B-V/T_{\rm eff}$,
we have obtained in this paper. This lends an additional support for
our $B-V/T_{\rm eff}$, and at the same time indicates that temperature
obtained with the Bell \& Gustafsson atmosphere suffers from 
errors for stars later than the G5 type ($T_{\rm eff}<5500$ K).
Flower's relation reflects this overestimate of temperature for
late type stars.

LCB adopts Flower's $B-V/T_{\rm eff}$ for $T_{\rm eff}>4250$ K
(log $T_{\rm eff}>3.63$), so that
it is identical with Flower's for this temperature range. Accordingly,
LCB's $B-V/T_{\rm eff}$ relation inherits the same problem as Flower's.

BL98 agree with ours for a rather wide range $3.68< \log T_{\rm eff}<3.78$
to within 0.02 mag. Beyond $3.78< \log T_{\rm eff}$ (6000 K) BL98 shows a
sudden break and turns away from our curve, giving significantly
bluer colour; At  log $T_{\rm eff}>3.80$ it is bluer by 0.04 mag than ours. 
B98 examined his surface brightness against IRFM of BL98 
for the main-sequence stars and found that BL98 give angular 
diameter larger by 4\% at log($T_{\rm eff}$)=3.95, while there is
no offset between the two at log($T_{\rm eff}$)=3.72. This 4\%
offset is consistent with BL98's $B-V$ bluer by 0.03-0.04 mag than ours.

Alonso et al (1996b)'s $B-V/T_{\rm eff}$ relation, which is based on IRFM with
ATLAS9 atmosphere supplemented with a calibration against angular diameter
measurements, is closely parallel to ours.  The agreement between
the two is $<0.01$ mag in $B-V$ for 3.75$<$log($T_{\rm eff}$)$<$3.85. The 
difference gradually increases as temperature decreases, and it becomes
0.03 mag at the end point of Alonso et al., log($T_{\rm eff}$)=3.70.

We have retained in Fig. 9 old Code et al. (1976)'s $B-V/T_{\rm eff}$
relation, which is based on a direct angular size measurement employing
intensity interferometry. Their curve smoothly matches with
ours with the difference is no more than 0.01 mag
in the overlapping range 3.76$<$log($T_{\rm eff}$)$<$3.85.

Our final assessment concerns the theoretical colour temperature relations
used by the Yale isochrone (Demarque et al. 1996)
and by Bertelli et al. (1994), and the one derived from ATLAS9.
The departure from
our empirical $B-V/T_{\rm eff}$ relation is significant, and the
difference can be 0.04 mag. The discrepancy is even larger with the
$B-V/T_{\rm eff}$ relation derived from the Kurucz (1991,
unpublished) atmosphere
(Bertelli et al. 1994): at  log($T_{\rm eff}$)=3.75, it gives 0.05 mag
redder than our relation. It is interesting to note that these
theoretical $B-V/T_{\rm eff}$ relations give values in agreement with
ours at log($T_{\rm eff}$)=3.65. This implies that the theoretical
relations, {\it if they are used to connect K5 stars with G5 stars},
raise systematic errors of 0.04 to 0.05 mag for relative colours
of stars between these two types. When translated into $M_V$, this
systematic offset amounts to 0.24-0.30 mag for a given $B-V$. 
We see the same trend with the calculation using ATLAS9 (Kurucz 1993)
atmosphere, 
although it shows larger deviations at both lower and higher 
temperatures from ours than the curve of 
Bertelli et al. who used 1991 version
of Kurucz' atmosphere.

\subsection{Giants}

A similar analysis is carried out for giants. We have plotted 
$\Delta(B-V)=(B-V)_0 - (B-V)_{\rm 0,Flower}$ in Fig. 11, taking this time
Flower's (1996) relation that covers the widest range as the zero point. The
figure includes classical work by Johnson (1966) and Ridgway et al. (1980),
which have been taken as the standard for long,
and recent work by LCB and BL98; a synthetic calculation using 
theoretical atmosphere of Kurucz (1993) is also included. 
As for Ridgway et al. (1980)'s data points, we assign $B-V$ colours 
from Hauck \& Mermilliod (1998) for the stars they used.
Since the scatter is to
large to obtain a sensible fit, however, we instead plot the points of
individual stars. We have also indicated the metallicity correction,
when [Fe/H] data are available, 
by arrows using the metallicity gradient given in eq. (2).

It is seen in Fig. 11 that Johnson (1966) and LCB are 0.03$-$0.04 mag
bluer than our $B-V/T_{\rm eff}$ relation for a $T=4000-4500$ K range,
and Flower (1996) are 0.03$-$0.04 redder in the same range. Unfortunately,
our formula does not reliably apply to the temperature lower than 
4000 K. BL98 give a $B-V/T_{\rm eff}$ relation with slope somewhat
steeper than ours, and the disagreement increases to $>$0.3 mag 
for $T<4300$ K ($\log T<3.63$). 
Ridgway et al.'s data are too noisy to make an accurate 
comparison, but it is likely that their colour temperature relation,
when transformed to the $B-V$ colour band, giving too blue colours, say by
0.10$-$0.15 mag for K giants (see also Flower 1996). 

A good agreement ($\Delta(B-V)<0.02$ mag) is seen between our curve and
Kurucz (1993) for a range 4200 K (the lowest temperature) and 5100 K.
Kurucz (1993) gives redder colours only for giants earlier than G type.

\section{Colour of the Sun}

$B-V$ colour of the sun has been playing an important role as a normalization
point for the stellar evolution models, yet observationally an
accurate measurement
of solar colours is notoriously difficult. Photometric observations 
yield 0.63 (Stebbins \& Kron 1957) to 0.69 (T\"ug \& Schmidt-Kaler 1982).
The method often adopted by observers is to use observations of
other stars, and interpolate and translate them to the sun, which typically
leads to 0.633$\pm$0.009 or 0.665$\pm$0.003 (Taylor 1998), or look for
``solar twins'' (de Strobel 1996, for a summary), rather than to
work directly with the sun.
For this solar analogue method to work properly, it is essential to
control the accuracy of temperature and also metallicity of these stars;
this is not easy a task, as we have seen in the preceding section.

In Fig. 9 the zero point at $T_{\rm eff}=
5777$ K is adjusted to LCB's value $B-V=0.633$. Our compilation shows that
all modern determinations
of the   $B-V/T_{\rm eff}$ relation give colours equal to or about 0.1 mag
bluer than this value. In particular, our newly obtained curve
gives $B-V=0.627$. The bluest value is given by Alonso et al. (1996b)'s
curve, which yields 0.621.  On the other hand, synthetic colors
from atmosphere models are significantly redder, 0.65 with the
colour-temperature relation  of the Yale isochrone, 
and 0.67 with Kurucz' atmosphere 
(Bertelli et al 1994;
Bessell, Castelli and Plez 1998).

Many solar analogue analyses in the past gave rather redder colour,
such as 0.66
(e.g., Hardrop 1978; Wamstecker 1981). de Strobel (1996) has argued that
Hardrop's sample is significantly metal rich, leading to redder
colour. This is also true with, e.g., Wamstecker's sample. The stars in
his sample have either luminosity lower than the sun or metallicity
higher than the sun by +0.15 dex. After careful selection de Strobel
concluded that $B-V=0.642\pm 0.004$.

Here we re-examine the case with de Strobel (1996)'s analysis in view of
our assessment for the temperature estimate.
She has given 26 stars on the list of effective-temperature-selected
solar analogues.
Among those stars 8 stars are given temperature by B98, and BL98's
temperature estimate is available for additional two stars (at the solar
temperature, BL98 and B98 agree very well). For 6 stars among these 
10 stars, de Strobel has given temperature based on spectroscopic 
studies much higher than BL98 and B98; the difference amounts to 100-160 K.
The average of the offset in 
the two $T_{\rm eff}$ estimates amounts to 63 K with de Strobel's 
temperature higher. The adjustment
of temperature can easily modify
de Strobel's estimate of $(B-V)_\odot$ into a bluer value 
by an amount of 0.02 or so. 

We have tried to find solar colour by fitting
the 8 stars with B98 data to the form 
$T_{\rm eff}=c_0+c_1(B-V)+f_1[{\rm Fe/H}]$
using the B98 temperature data.
In spite of a small sample and a small range, 
the fit is well constrained, yielding
$(B-V)_\odot=0.61$, with metallicity gradient $\partial T_{\rm eff}/
\partial {\rm [Fe/H]}=$ 220 K/dex and temperature gradient of the right
order of magnitude, although this temperature-selected sample is clearly
too narrow in the temperature range to find the correct gradient
(see Fig. 12a).
When we replace the B98 temperature with
de Strobel's ([Fe/H] data are not replaced), however, we obtain
the temperature basically constant at 5820 K and metallicity
gradient with the sign opposite to what we have obtained with the B98
temperature (Fig. 12b).  This indicates that the spectroscopic temperature
does not agree with the bolometry-surface brightness estimate
we used in this paper.
This uncertainty in temperature estimations tells us a difficulty of 
an accurate estimate of 
solar colour from solar analogues: one must know star's
temperature to an absolute accuracy of $\pm 50$ K and [Fe/H]$<$ 0.1.
The former is the accuracy one can barely achievable with the 
bolometry-surface brightness estimate as we have seen in this paper.

From our $B-V/T_{\rm eff}$ relation we
conclude $(B-V)_\odot=0.627\pm0.014\pm0.012$
with the two errors standing for uncertainty of the estimate of the
locus of the relation and possible intrinsic dispersion for the sun
around the relation. All modern $B-V/T_{\rm eff}$ relations (other than
synthetic) documented
in Fig. 9 give $(B-V)_\odot$ within this error.
We are not able to reconcile our value
with a red colour 0.66-0.67, often referred to in the literature.
If the sun would really be this red, it is significantly off from normal
G2 stars.

Our final remark concerns colour of the sun from spectroscopic synthesis that
synthesis calculations using the measured solar spectrum
tend to give bluer colours 0.61-0.65
(see Fukugita, Ichikawa \& Sekiguchi 1999 for details), and this agrees 
with the value we have inferred above.

\section{Metallicity dependence}
\label{sec_feh_dep}

We present the metallicity dependence of our  $B-V/T_{\rm eff}$ relation
in Fig. 13. Three solid curves represent the relations for
[Fe/H]=$-$1.5, $-$0.5, 0 and 0.3 for main sequence.  
We limit our plot to $0.3<B-V<1.0$,
for which our metallicity dependence is well constrained by data.
We overlay the curves of Alonso et al. (1996b) and the one we
computed with Kurucz (1993) atmosphere, where Kurucz' temperature
is scaled down by 200 K to make figure ease a comparison  
in the same figure. 

With our curve $\partial T_{\rm eff}/\partial {\rm [Fe/H]}=325\pm20$
K/dex, owing to the absence of a [Fe/H]*$(B-V)$ term as discussed
in section 2.5.  This means for G2 stars 
$\partial (B-V)/\partial [{\rm Fe/H}]= 0.9$ mag/dex for the portion contributed
from the atmosphere.
If we adopt the fit where the cross term is taken
as a free parameter the partial derivative increases towards bluer
side by 50 degree between $B-V=0.8$ and 0.4. 

Apparently a good gross agreement is seen between  Alonso et al. and ours.
A more careful examination, however, shows that metallicity gradient
of the former increases from 330 K/dex at $B-V=0.8$ to 480 K/dex at 0.4. 
We have not seen this large change with our fit, even if we allow for
the cross term: our analysis shows that
it is at most one third this value.

The family of curves calculated with Kurucz (1993)'s
atmosphere is  generally shifted by about
200 K  to a lower temperature. With shifting by this amount, the
curve derived from Kurucz's atmosphere agrees well with ours for the
range $0.5<B-V<0.7$. On the other hand, the metallicity gradient
shows a very good agreement with ours: 
 $\partial (B-V)/\partial {\rm [Fe/H]}$
stays between 350$-$370 K/dex in the range $B-V=0.4-0.8$
with the Kurucz atmosphere.
The metal dependence is well accounted for with the Kurucz atmosphere
for F-K stars.

\section{Conclusion}

In this paper we have used the modern, least model-dependent determination of 
effective temperature of stars and the best available data for 
$B-V$ colours,
metallicity and surface gravity in order to derive a $B-V$ colour 
temperature relation with metallicity and gravity as auxiliary parameters. 
We have achieved the smallest residual temperature over those available
in the literature. The fit
we obtained is well constrained and a number of tests assure the quality
of data and show a consistency
of the data processing procedures. Our relation covers the range 
from F0 to K5 stars ($T_{\rm eff}=4000-7000$ K)
and [Fe/H] from $-$1.5 to +0.3 both for dwarfs and giants. The dispersion
of the fit, 66 K, perhaps represents the limit we can achieve
with the present quality of data. 

The most important limiting factors are temperature determinations
and metallicity measurements. This means that we can attain accuracy of
0.02 mag in $B-V$ colours for a given temperature for F0-K0 stars, and
slightly worse for later type stars. This is still a significant error,
but it is not as large as the disagreement recognized 
among various isochrone works.

We have examined various  $B-V/T_{\rm eff}$ relations available in the
literature. Our relation smoothly joins the Code et al. (1976)'s relation
given for high temperature stars, and also
shows a close match with Alonso et al. (1996b)'s
$B-V/T_{\rm eff}$ relation which is based on the IRFM with additional
calibrations. BL98 give
a relation that agrees reasonably well with ours, but a significant
discrepancy is observed for stars earlier than F8 stars. The relations
of Flower (1996) and LCB are largely off from ours, as much as
0.1 mag for low temperature stars. Colour becomes significantly redder
for G5 or later type stars, if these $B-V/T_{\rm eff}$ relations are
used. Our comparison demonstrates that some calibration of temperatures
against those obtained from angular diameter measurements consists an
essential element for a high accuracy. 

We also clarified
systematic errors with the colour-temperature relation
obtained by synthetic computation using model atmospheres: they 
deviate significantly from our empirical relation. 
In particular, the offset changes 
by 0.04-0.05 in $B-V$ across K0 to K5 stars, which directly induces an
error in the slope of the colour-magnitude diagram by this amount. 
This means that, for
instance, if the distance to one open cluster is calibrated with K5 
stars and if another is with K0 or some earlier stars, we would be
led to an error of 10-15\% in distance. This error also makes intermediate age
galaxies, which contains G stars as a major source of the bluer component, 
appreciably redder in stellar population synthesis model of galaxies.

Another implication important for cosmology is that the population 
synthesis model of Bruzual \& Charlot
(1993) (see also Charlot, Worthey \& Bressan 1996) would give too blue 
colour for early type galaxies in their late stage of evolution. Typically
5 Gyr after the initial burst, G and K giants start dominating the
light from elliptical galaxies, and Bruzual \& Charlot take Ridgway et
al (1980)'s temperature scale to assign colours to tracks. We have shown 
that Ridgway et al.'s scale gives 
typically 0.1 mag bluer at a given temperature
for K giants. Therefore, we should make Bruzual \& Charlot's  $B-V$ colour 
prediction redder by this amount. With this revision, the burst model
would give $B-V=0.95$ already at 5 Gyr from the burst, rather than
9 Gyr in their original model.  This offset also explains the discrepancy,
at least in part,
between the predictions of Bruzual \& Charlot and of Bertelli et al. (1994),
the latter, using the Kurucz atmosphere, giving 0.05 magnitude redder than
the former.      

We have also studied the problem of colour of the sun. Our   
$B-V/T_{\rm eff}$ relation gives $(B-V)_\odot=0.63\pm0.02$ which
agrees with ``long wavelength group'',
but disagree with ``short wavelength group'' of solar colour
(Taylor 1998).  We have emphasized the importance to accurately
estimate temperature
and metallicity when colour of the sun is inferred from solar
analogues. The quality of  temperature
and metallicity determinations of the presently available sample 
is probably insufficient 
to determine colour of the sun within an error of 0.02 mag.

\acknowledgements

We are grateful to Drs. R. L. Kurucz, J.-C. Mermilliod and S. Yi 
for kindly providing us with their data in machine readable form.
One of us (MF) thanks the Raymond and Beverly Sackler Fellowship and the
Alfred P. Sloan Foundation for the support for the work in
Princeton.

\newpage 
\begin{deluxetable}{lcc}
\tablewidth{0pc}
\tablecaption{Colour temperature relation and derivatives\label{table1}}
\tablehead{\colhead{$B-V$} &\colhead{$T_{\rm eff}$} 
& \colhead{${\partial T_{eff}\over \partial(B-V)}\times 10^{-2}$}}
\startdata
0.3 & 7078 & $-$48 \cr
0.4 & 6621 & $-$43 \cr
0.5 & 6214 & $-$38 \cr
0.6 & 5852 & $-$34 \cr
0.7 & 5532 & $-$30 \cr
0.8 & 5248 & $-$27 \cr
0.9 & 4996 & $-$24 \cr
1.0 & 4770 & $-$21 \cr
1.1 & 4567 & $-$19 \cr
1.2 & 4382 & $-$18 \cr
1.3 & 4209 & $-$17 \cr
1.4 & 4044 & $-$16 \cr
1.5 & 3882 & $-$16 \cr
\enddata
\end{deluxetable}

\newpage 
\begin{deluxetable}{rcccccccc}
\scriptsize
\tablewidth{0pc}
\tablecaption{Data for the temperature check\label{table2}}
\tablehead{ \colhead{BS} & \colhead{LC} & \colhead{$\phi$} & \colhead{$(V-K)_{0}$} &
\colhead{$F_{\rm bol}$} & \colhead{ref/norm} & \colhead{$T_{\rm eff}$(ref)} &
\colhead{$\Delta T_{\rm eff}$(B98)} & \colhead{$T_{\rm eff}$(BL98)} }
\startdata
 & & (mas) & & {\scriptsize ($10^{-6}$erg cm$^{-2}$s$^{-1}$) } &  & (K) & (K) & (K)\cr
  2491&  V& 5.92$\pm$.09&-.099 &114.3$\pm$4.4 & a/A &9947$\pm$122&$-$3&\cr
  7001&  V& 3.24$\pm$.07&$-$.001& 30.4$\pm$1.2 & a/A &9655$\pm$141&$-$184&\cr
  7001&  V& 3.24$\pm$.07&$-$.001& 29.0$\pm$2.6 & b/B &9542$\pm$237&$-$71&\cr
  6879&  V& 1.44$\pm$.06&.047& 5.53$\pm$0.22 & a/A &9458$\pm$218&$-$203&\cr
  4534&  V& 1.33$\pm$.10&.140& 3.61$\pm$0.13 & a/A &8847$\pm$342&16&\cr
  8728&  V& 2.10$\pm$.14&.144& 8.80$\pm$0.31 & a/A&8797$\pm$303&50 & 8622$\pm$86\cr
  6556&  III& 1.63$\pm$.13&.379& 3.65$\pm$0.13 & a/A &8013$\pm$327&$-$16 & 7883$\pm$63\cr
  6556&  III& 1.63$\pm$.13&.379& 3.640$\pm$0.218 & b/B &8008$\pm$341&$-$10 & 7883$\pm$63\cr
  2943&  IV.5&5.51$\pm$.05&1.010& 18.08$\pm$0.76 & a/A &6502$\pm$74&56&\cr
  2943&  IV.5&5.51$\pm$.05&1.010& 18.35$\pm$0.367 & c/C&6526$\pm$44&32&\cr
   Sun&  V  &19193E2&1.486& 1.370(2)E+12 & d/D &5780$\pm$2&$-$16&\cr
  2990&  III&8.04$\pm$.08&2.246& 11.82$\pm$0.47 & e/E &4840$\pm$54&$-$45 & 4837$\pm$29 \cr
  2990&  III&8.04$\pm$.08&2.246& 11.78$\pm$0.12 & f/A+B &4836$\pm$27&$-$41 & 4837$\pm$29 \cr
  2990&  III&8.04$\pm$.08&2.246& 11.69$\pm$0.47 & g/F &4827$\pm$54&$-$31 & 4837$\pm$29 \cr
  7949&  III&4.62$\pm$.04&2.398& 3.598$\pm$0.036 & f/A+B &4743$\pm$24&$-$88 & 4732$\pm$28\cr
   168&  III&5.64$\pm$.05&2.430& 5.00$\pm$0.15 & h/ &4660$\pm$41&$-$34&\cr
   617&  III&6.85$\pm$.06&2.630& 6.43$\pm$0.20 & h/ &4503$\pm$40&$-$42&\cr
   165&  III&4.12$\pm$.04&2.853& 2.105$\pm$0.021 & f/A+B &4392$\pm$24&$-$93 & 4335$\pm$30\cr
  5340&  III&20.95$\pm$.20&2.921& 49.21$\pm$0.49 & f/A+B &4283$\pm$23&$-$28&\cr
  5340&  III&21.00$\pm$.20&2.921& 49.21$\pm$0.49 & f/A+B &4278$\pm$23&$-$23&\cr
  5340&  III&20.95$\pm$.20&2.921& 49.77$\pm$0.10 & g/F &4295$\pm$30&$-$41&\cr
  5340&  III&21.00$\pm$.20&2.921& 49.77$\pm$0.10 & g/F &4290$\pm$30&$-$35&\cr
  6705&  III&10.13$\pm$.24&3.503& 8.408$\pm$0.084 & f/A+B &3960$\pm$48&$-$20&\cr
  6705&  III&10.20$\pm$.20&3.503& 8.408$\pm$0.084 & f/A+B &3946$\pm$40&$-$6&\cr
  6705&  III&10.13$\pm$.24&3.503& 8.330$\pm$0.417 & b/B &3951$\pm$68&$-$10&\cr
  6705&  III&10.20$\pm$.20&3.503& 8.330$\pm$0.417 & b/B &3937$\pm$63&3&\cr
  6705&  III&10.13$\pm$.24&3.503& 8.59$\pm$0.34 & g/F &3981$\pm$61&$-$41&\cr
  6705&  III&10.20$\pm$.20&3.503& 8.59$\pm$0.34 & g/F &3968$\pm$55&$-$27&\cr
  1457&  III&20.88$\pm$.10&3.704& 33.49$\pm$0.335 & f/A+B &3897$\pm$13&$-$37&\cr
  1457&  III&21.21$\pm$.21&3.704& 33.49$\pm$0.335 & f/A+B &3866$\pm$21&$-$7&\cr
  1457&  III&20.88$\pm$.10&3.704& 33.82$\pm$1.35 & g/F &3906$\pm$40&$-$47&\cr
  1457&  III&21.21$\pm$.21&3.704& 33.82$\pm$1.35 & g/F &3876$\pm$43&$-$16&\cr
\tablerefs{
 (a) Code et al. (1976), (b) Leggett et al. (1986), (c) Alosno et al. (1995), 
 (d) Bahcall \& Pinsonneault (1995), (e) Blackwell \& Lynas-Grey (1994), 
(f) Blackwell et al. (1990),
 (g) Di Benedetto \& Rabbia (1987), (h) Faucherre et al. (1983),
(A) $3.36\times10^{-9}$ (erg cm$^{-2}$s$^{-1}$) at 5556\AA~~ Oke \& Schild (1970); 
(B) $3.39\times10^{-9}$ (erg cm$^{-2}$s$^{-1}$) Hayes \& Latham (1975)
(C) $3.47\times10^{-9}$ (erg cm$^{-2}$s$^{-1}$) T\"ug, White \& Lockwood (1977);
(D) ;
(E) $3.45\times10^{-9}$ (erg cm$^{-2}$s$^{-1}$) Dreiling \& Bell (1980); 
(F) $3.42\times10^{-9}$ (erg cm$^{-2}$s$^{-1}$) Hayes (1979)
}
\enddata
\end{deluxetable}

\newpage 
\begin{deluxetable}{lcc}
\tablewidth{0pc}
\tablecaption{Fit parameters\label{table3}}
\tablehead{ &\colhead{full sample}&\colhead{main sequence}}
\startdata
range & 0.3$<B-V_0<$1.5 &  0.3$<B-V_0<$0.9 \cr 
 \tableline
data points & 266 &  104 \cr
 \tableline
$c_0$ &  3.98319 $\pm$ 0.00727 &  3.97542 $\pm$  0.01441 \cr
$c_1$ & $-$0.42998 $\pm$ 0.02630 & $-$0.40671 $\pm$ 0.06180 \cr
$c_2$ &  0.18174 $\pm$ 0.02999 &  0.16881 $\pm$ 0.08292 \cr
$c_3$ & $-$0.04280 $\pm$ 0.01050 & $-$0.04830 $\pm$ 0.03413\cr
$f_1$ &  0.02691 $\pm$ 0.00173 &  0.02576 $\pm$ 0.00234 \cr
$f_2$ &  0.00437 $\pm$ 0.00100 &  0.00454 $\pm$ 0.00128 \cr
$g_1$ & $-$0.003223 $\pm$ 0.000648 &   $-$0.003223 (fixed) \cr
\enddata
\end{deluxetable}

\newpage 
\begin{deluxetable}{llccc}
\tablewidth{0pc}
\tablecaption{Summary of estimated errors (dispersion)\label{table4}}
\tablehead{ \colhead{source} & \colhead{error} & \colhead{$\Delta T_{\rm eff}$(F5)} &
             \colhead{$\Delta T_{\rm eff}$(G2)} & \colhead{$\Delta T_{\rm eff}$(K5)} }
\startdata
$T$             &        & 6500 & 5800 & 4500 \cr
\tableline
$\Delta T$      &        & 40 & 40 & 40\cr
$\Delta (B-V)$  & 0.01   & 41 & 33 & 18\cr
$\Delta E(B-V)$ & 0.003 &  14 & 11 &  6\cr
$\Delta {\rm [Fe/H]}$ & 0.15   & 50 & 50 & 50\cr
$\Delta \log g$ & 0.2    &  6 &  6 & 6 \cr
\tableline
 quadrature sum &        & 77 & 73 & 67 \cr
\tableline
$\sigma$ of the fit & & 80 & 66 & 60 \cr
\enddata
\end{deluxetable}

\newpage 
\begin{deluxetable}{ll}
\tablewidth{0pc}
\tablecaption{Literature appeared in Figs. 8 and 9 \label{table5}}
\tablehead{ \colhead{legend} &\colhead{reference : method : range} }
\startdata
 Code76   & Code et al. 1976 : $\phi$ (interferometry) and $F_{\rm bol}$ : $T_{\rm eff}>5780K$, LC=V\cr
 VB85     & VandenBerg \& Bell 1985 : Bell \& Gustafsson 1978 atmosphere \cr
 Bertelli94 & Bertelli et al. 1994 : Kurucz 1993 atmosphere for $T_{\rm eff}>4000$K \cr
 Yale96   & Demarque et al. 1996 : model atmosphere \cr
 Flower96 & Flower 1996 : compilation of $T_{\rm eff}$ : LC=V-III and II \cr
 Alonso96 & Alonso et al. 1996 : IRFM (Kurucz 1991, 1993) : $0.2<B-V<1.5$, 
LC=V\cr
 LCB   & Lejeune et al. 1997 : used Flower 1996 for $T_{\rm eff}>4250$K: \cr
       & \hskip20mm  $\log g$ and [Fe/H] extension with Kurucz atmosphere\cr
 BL98 & Blackwell \& Lynas-Gray 1998 : IRFM (Kurucz 1992): 
             4000$<T_{\rm eff}<$9000K, LC=V,IV,III\cr
\enddata
\end{deluxetable}


\newpage

\begin{figure}
\epsscale{0.9}
\plotone{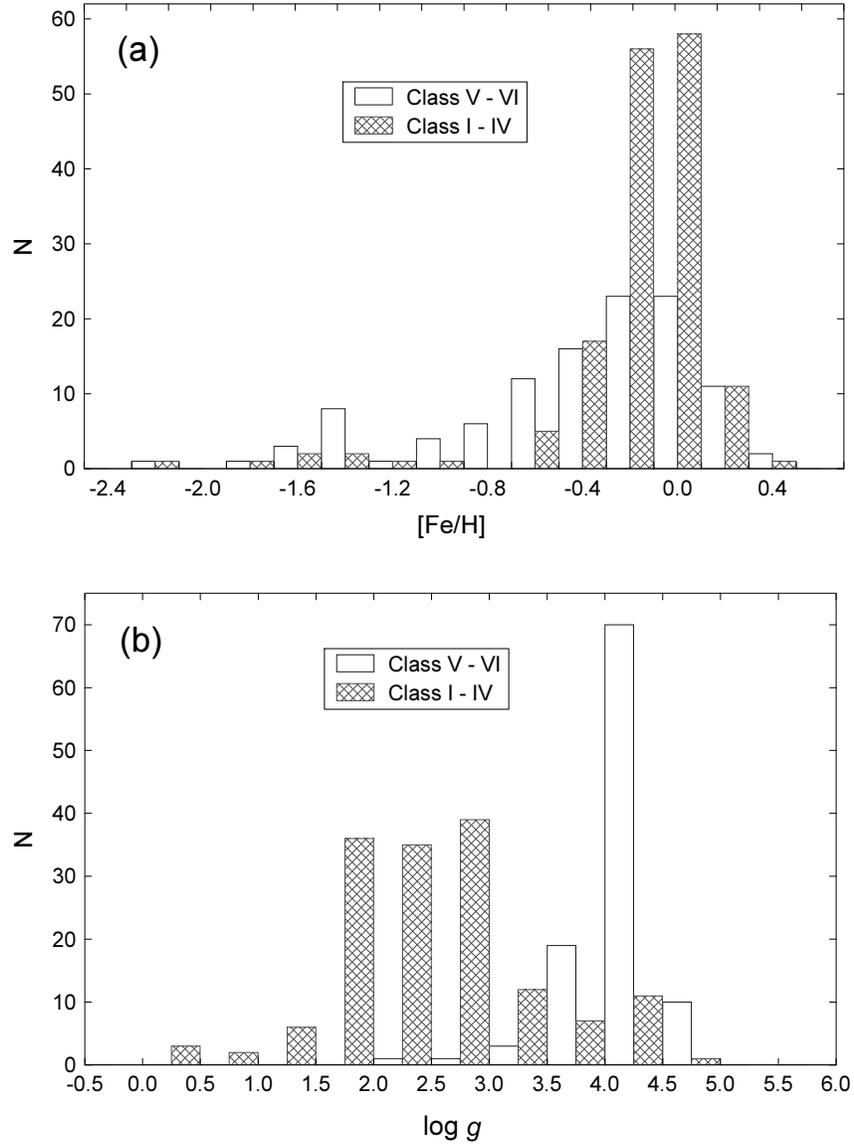}
\figcaption
{Distribution of (a) [Fe/H] and (b) $\log g$ for the 270 stars used in 
our analysis.}
\end{figure}

\begin{figure}
\epsscale{0.9}
\plotone{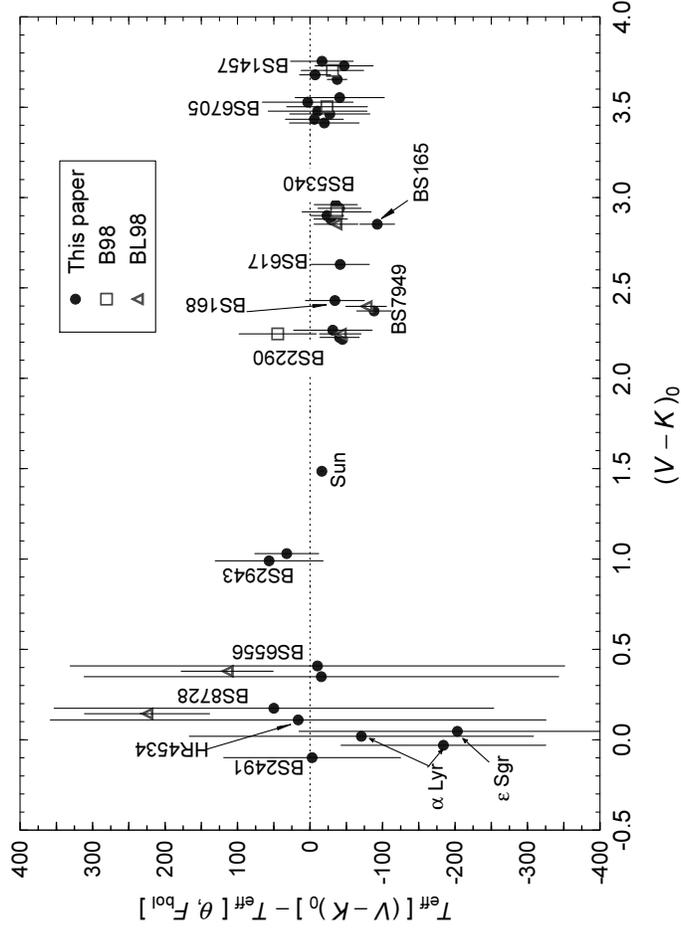}
\figcaption
{Examination of temperature given by B98. The difference $\Delta T_{\rm eff}=
T_{\rm eff}[(V-K)_0]-T_{\rm eff}({\rm reference})$ (B98) 
is plotted as a function of $(V-K)_0$. The data points used by B98 for his
own examination and
temperature given in BL98 are also plotted.
The data actually used in our analysis is confined to
$(V-K)_0\ge1.0$.}
\end{figure}

\begin{figure}
\epsscale{0.9}
\plotone{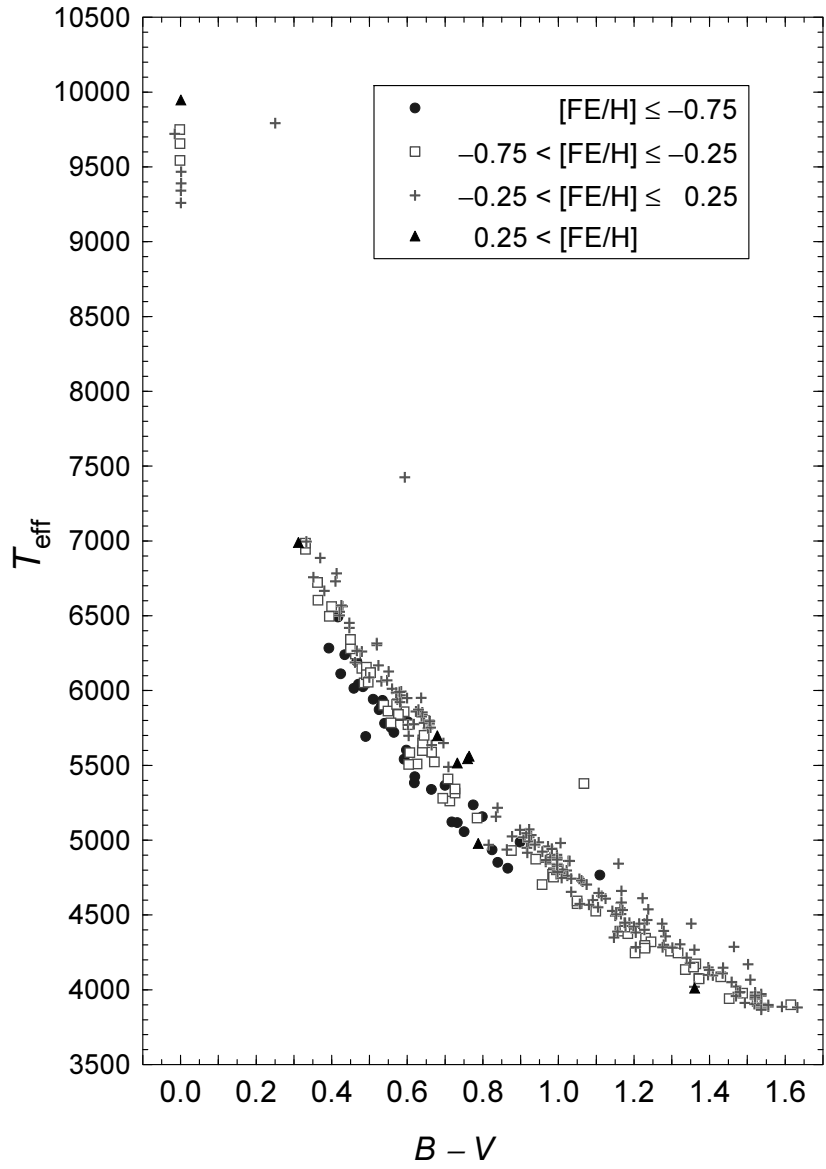}
\figcaption
{$T_{\rm eff}$ as a function of $B-V$ for the full sample. Different symbols
are used to represent metallicity grouped into 4 classes.}
\end{figure}

\begin{figure}
\epsscale{0.9}
\plotone{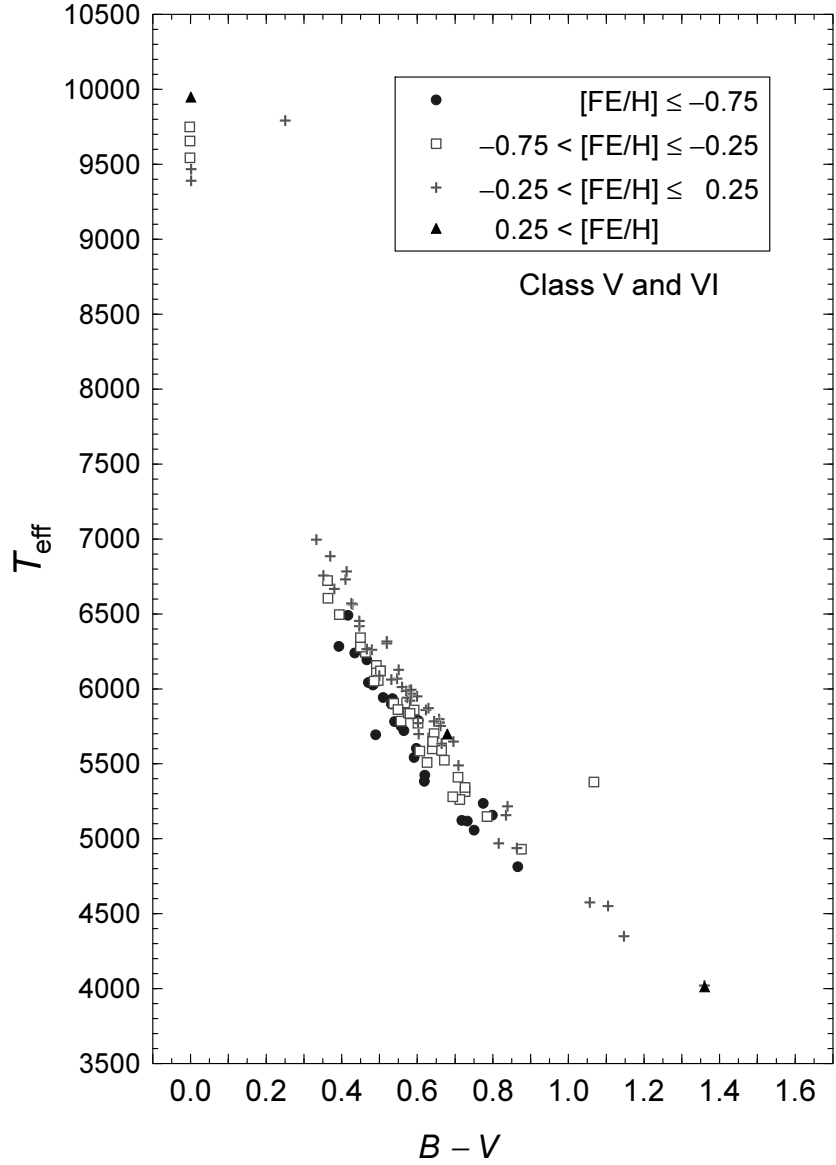}
\figcaption{Same as Figure 3, but for dwarfs.}
\end{figure}

\begin{figure}
\epsscale{0.9}
\plotone{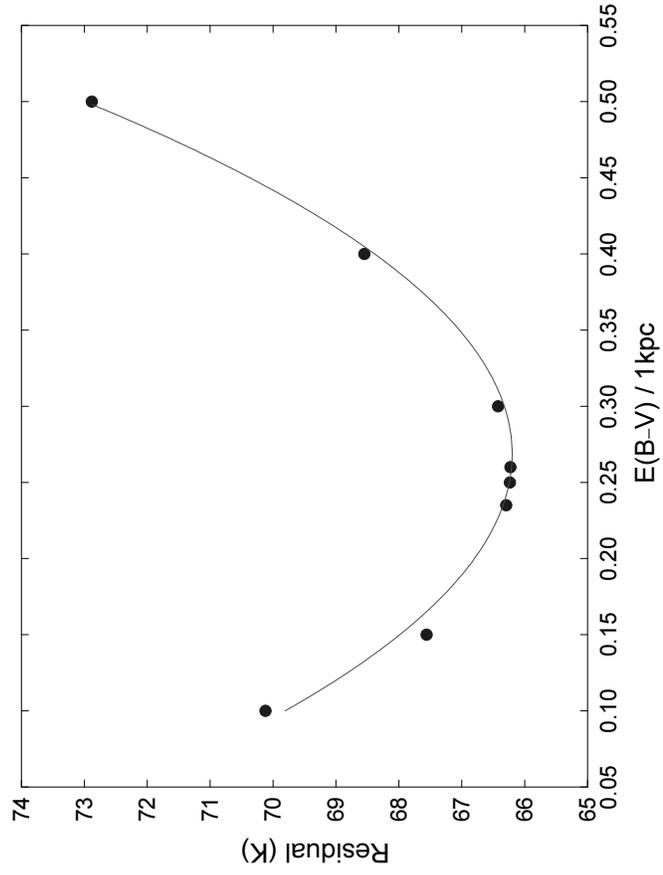}
\figcaption
{Residual temperature of the fit as a function of the selective extinction
per unit distance. The curve is an interpolation.}
\end{figure}

\begin{figure}
\epsscale{0.9}
\plotone{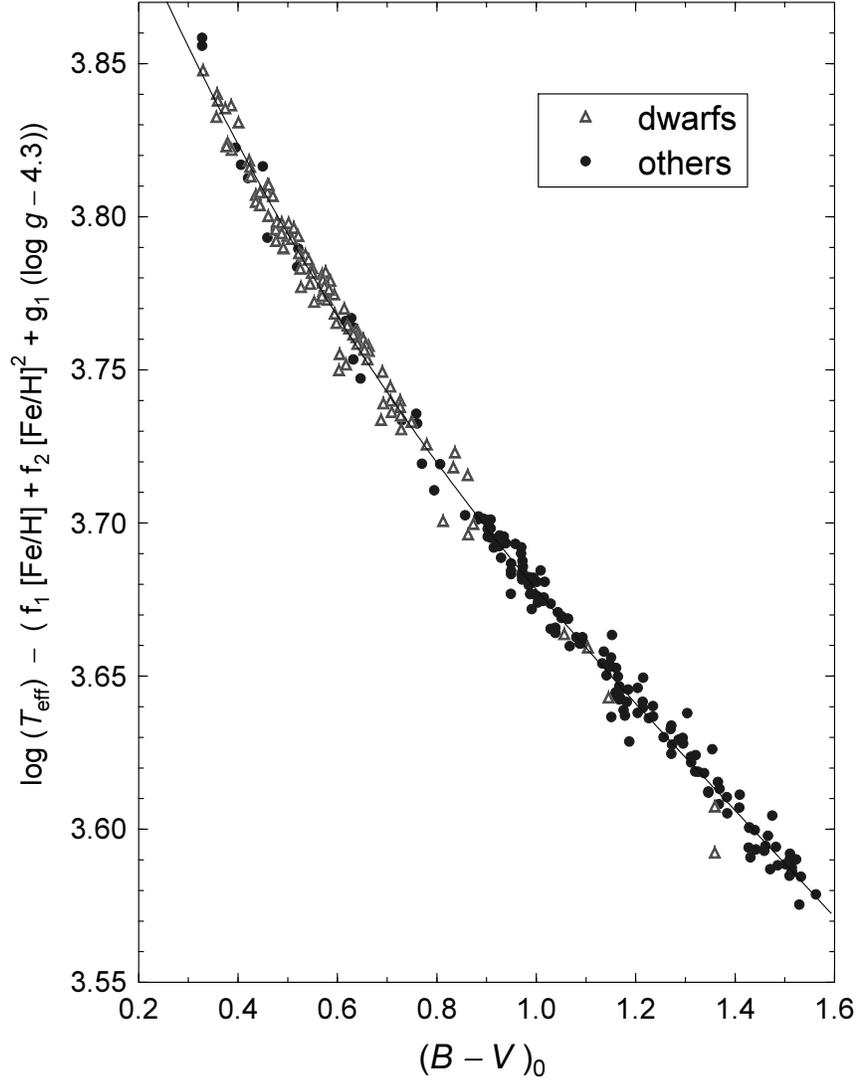}
\figcaption
{Quality of the fit. The deviation of temperature from the fit is
shown for each stars after subtracting expected metallicity 
contribution and scaled to
$\log g=4.3$. Open triangles 
represent dwarfs and solid circles giants. }
\end{figure}
 
\begin{figure}
\epsscale{0.9}
\plotone{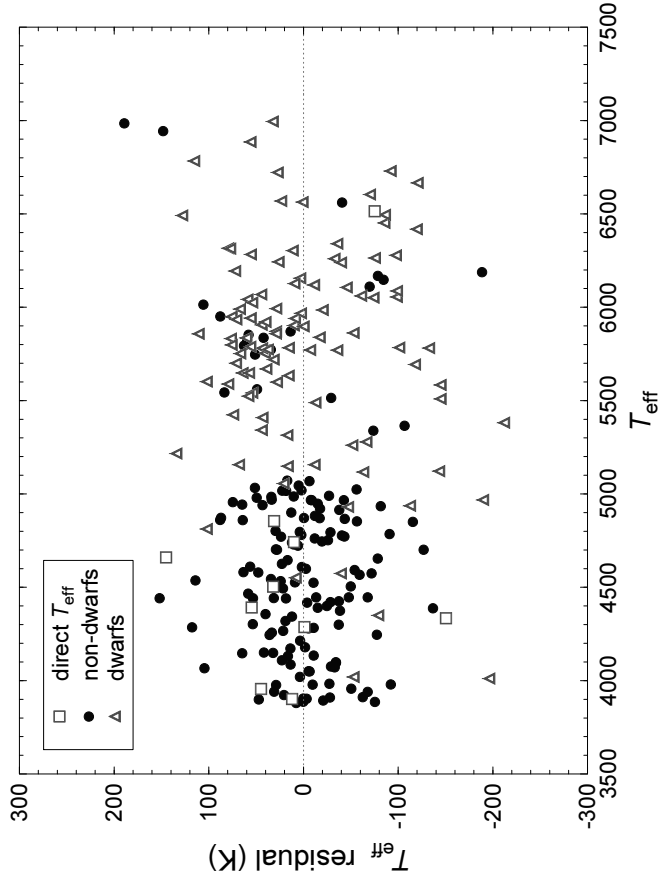}
\figcaption
{Residual of the fit derived from Fig. 6. Open triangles 
represent dwarfs and solid circles giants. 
The stars for which direct 
temperature measurement are available are plotted with open squares
(median value of the data shown in Fig. 2).}
\end{figure}

\begin{figure}
\epsscale{0.9}
\plotone{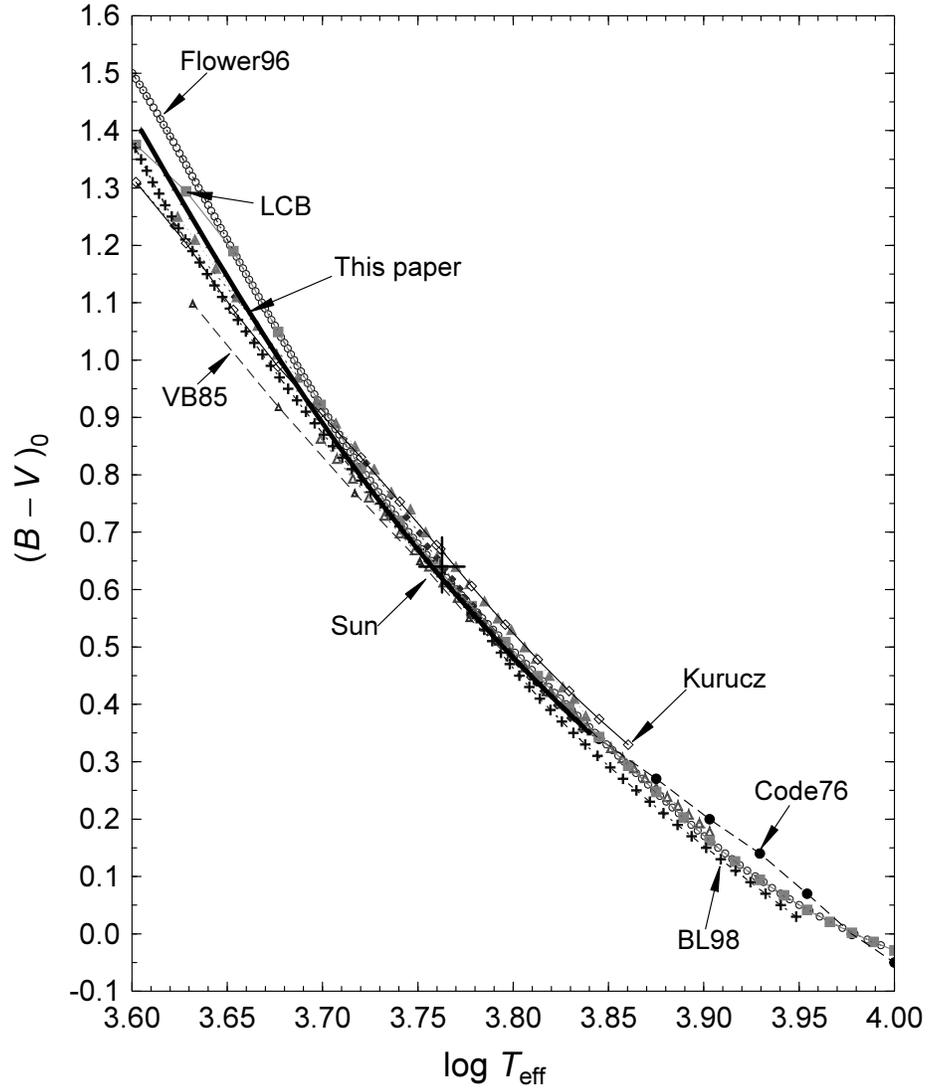}
\figcaption
{Compilation of $B-V$ colour temperature relations for dwarfs
available in the literature, as compared with the one obtained in this
paper.  The relation is given for the solar abundance.}
\end{figure}

\begin{figure}
\epsscale{0.9}
\plotone{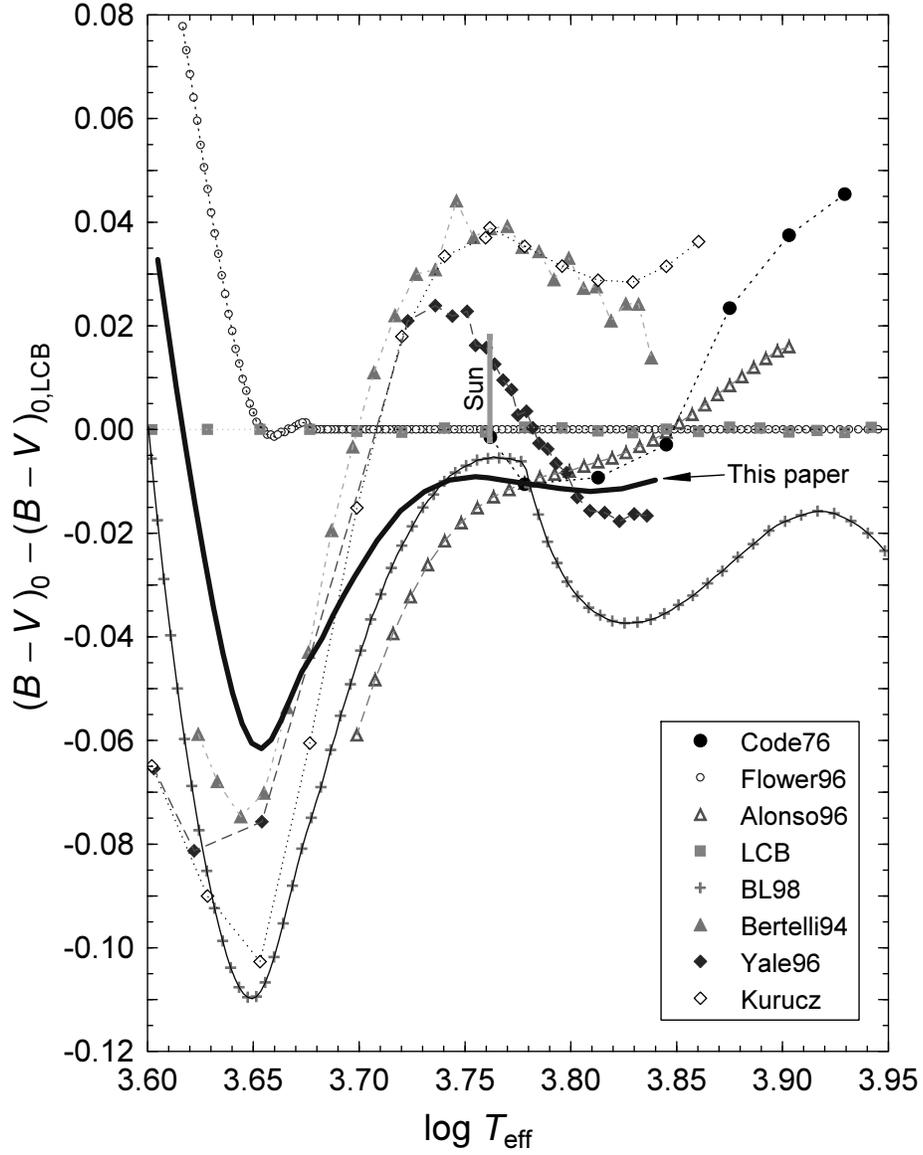}
\figcaption
{Comparison of $B-V$ colour temperature relations for dwarfs.
The difference is plotted for various $B-V$ colour temperature relations,
taking the one by LCB, which covers the widest parameter range 
as the zero point. All relations are plotted at solar metallicity.
The position of the sun with $B-V=0.64\pm0.01$ and $T_{\rm eff}$=5777 K
is also shown. For legends see Table 5.}
\end{figure}

\begin{figure}
\epsscale{0.9}
\plotone{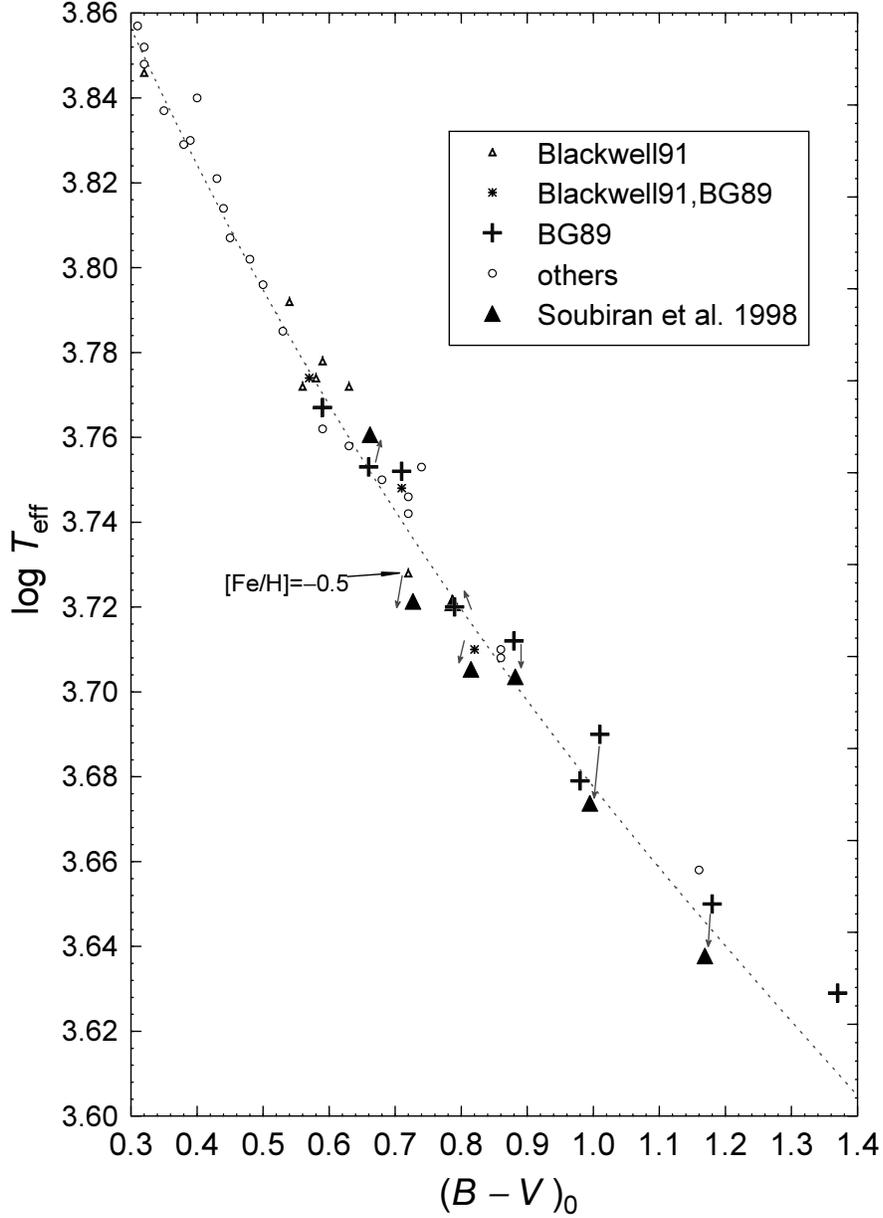}
\figcaption
{The data points for $B-V$ colour temperature relation used by 
Flower (1996), overlayed with new determination of the temperature
for the same stars (when available), the change indicated by arrows. 
One point with given metallicity is star largely deviated from
solar.  The dotted curve is the $B-V$ colour temperature relation
given in eq. (2). BG89 stands for Bell \& Gustafsson (1989).}
\end{figure}

\begin{figure}
\epsscale{0.9}
\plotone{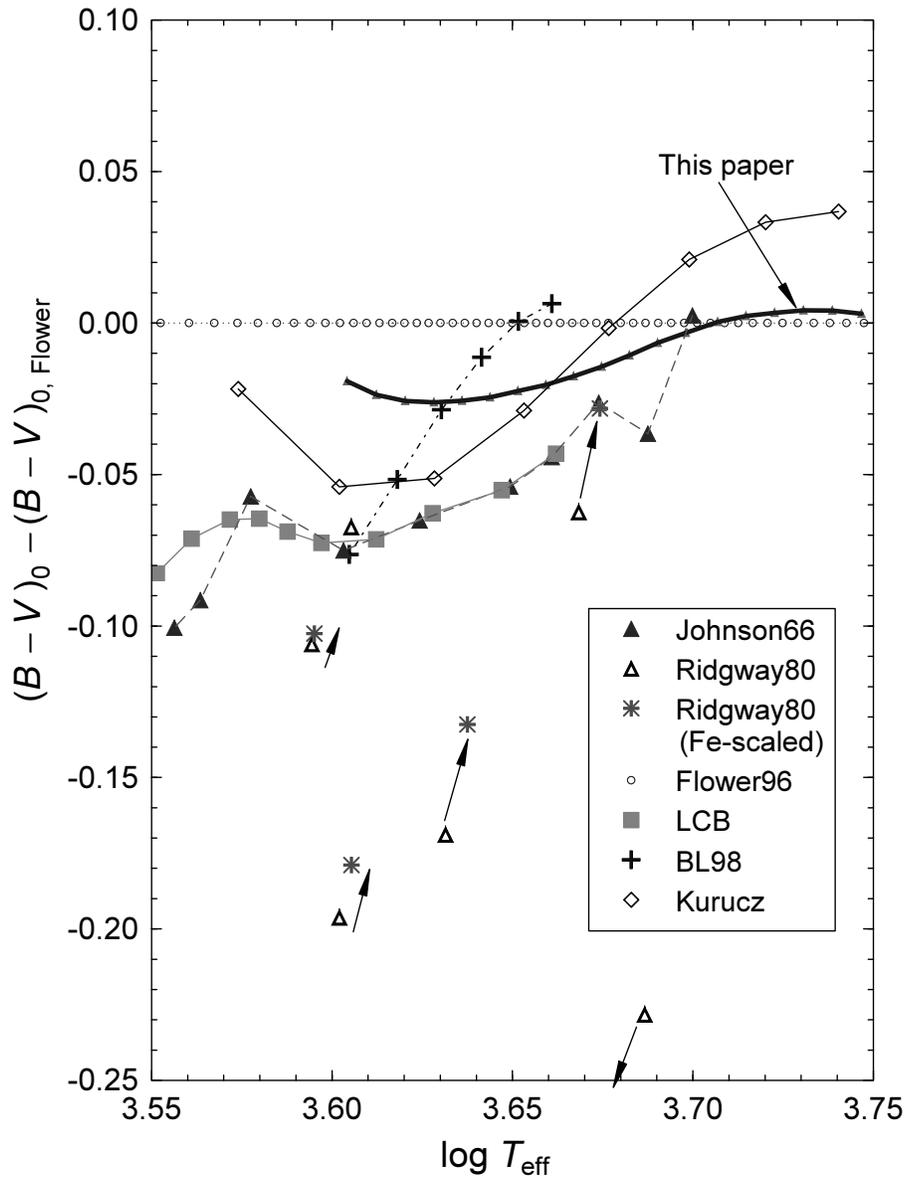}
\figcaption
{Comparison of the $B-V$ colour temperature relations for giants.
The relation of Flower (1996), which covers the widest range, is 
taken as the zero point. The arrows show the metallicity correction
applied to the data points by Ridgway et al. (1980).}
\end{figure}

\begin{figure}
\epsscale{0.9}
\plotone{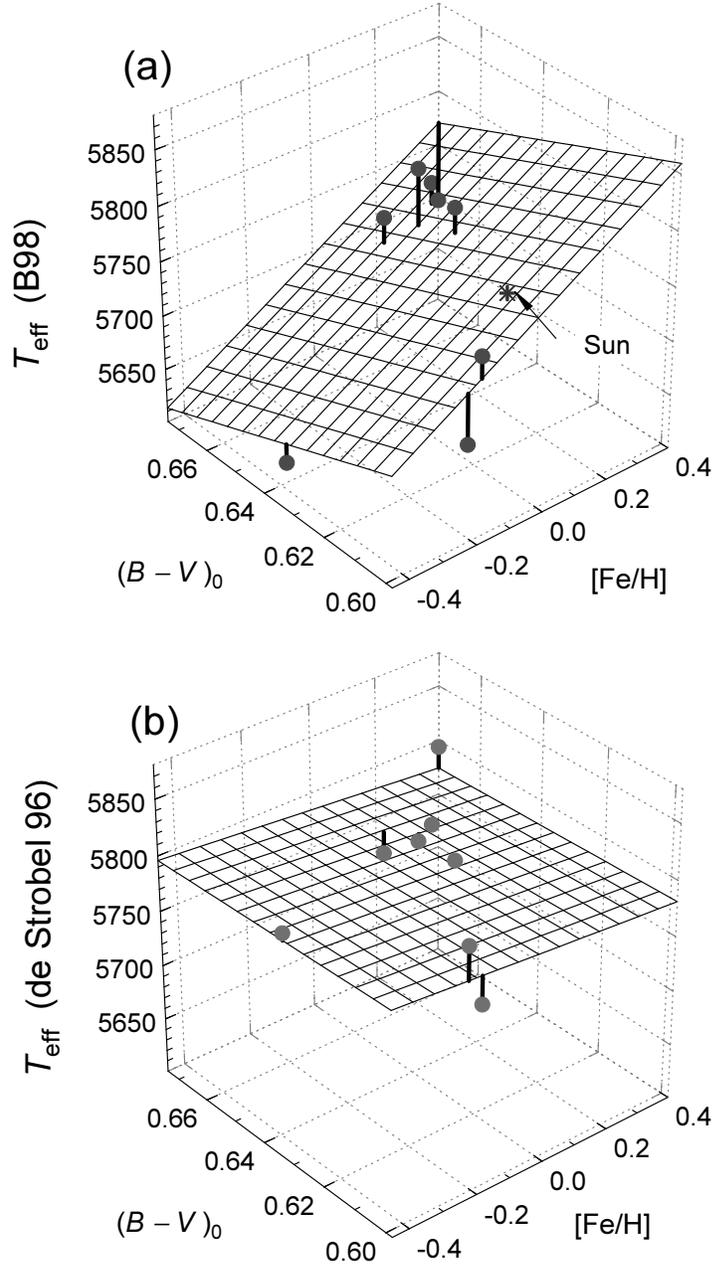}
\figcaption
{(a) Solar analogue stars of de Strobel (1996) selected on the basis
of temperature in $T_{\rm eff}$, $(B-V)_0$, [Fe/H] space. Only those 
stars that have B98 temperature estimates
are plotted, and temperature is taken from B98. The plane shows a
bilinear fit, and line segment attached to data are distances
from the plane. (b) Same as (a) but with the temperature given by
de Strobel.}
\end{figure}

\begin{figure}
\epsscale{0.9}
\plotone{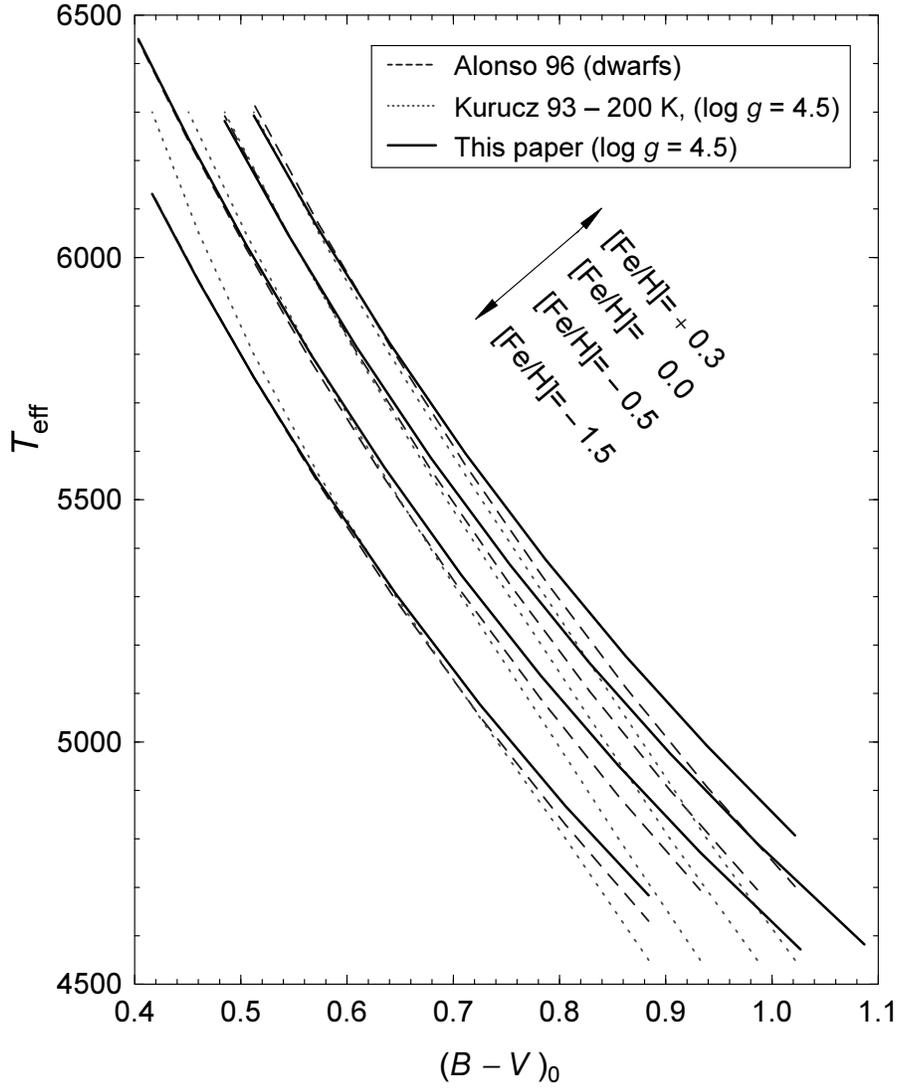}
\figcaption
{Metallicity dependence of the $B-V$ colour temperature relation for
dwarfs. The curves are plotted for [Fe/H]=$-1.5, -0.5, 0$ and +0.3.
The curves derived from Kurucz atmosphere are shifted downward by 200 K 
to ease a comparison.}
\end{figure}

\end{document}